\documentclass{IEEEtran}
\usepackage{cite}
\usepackage{amsmath,amssymb,amsfonts}
\usepackage{graphicx}
\usepackage{textcomp,nicefrac}
\usepackage{amssymb}
\usepackage{subfigure}
\usepackage{siunitx}
\usepackage[switch]{lineno}
\usepackage{xcolor}
\usepackage{multicol}
\usepackage{multirow}
\usepackage{url}
\usepackage[colorlinks,
            linkcolor=blue,       
            anchorcolor=blue,  
            citecolor=blue]{hyperref}

\def\BibTeX{{\rm B\kern-.05em{\sc i\kern-.025em b}\kern-.08em
T\kern-.1667em\lower.7ex\hbox{E}\kern-.125emX}}
\begin{document}
\title{The performance of large-pitch AC-LGAD with different N+ dose}

\author{Mengzhao Li, Weiyi Sun, Zhijun Liang, Mei Zhao, Xiaoxu Zhang, Yuan Feng, Yunyun Fan, Tianya Wu, Wei Wang, Xuan Yang, Bo Liu, Shuqi Li, Chengjun Yu, Xinhui Huang, Yuekun Heng, \IEEEmembership{Member, IEEE}, Gaobo Xu

\thanks{This work was supported in part by the National Natural Science Foundation of China under Grant 12042507, Grant 12175252, Grant 12275290, Grant 11961141014, Grant 12105298; in part by the China Postdoctoral Science Foundation under Grant 2022M722964; in part by the National Key R$\&$D Program of China under Grant 2022YFE0116900; in part by the State Key Laboratory of Particle Detection and Electronics under Grant SKLPDE-ZZ-202214, Grant SKLPDE-ZZ-202204; and in part by the Scientific Instrument Developing Project of the Chinese Academy of Sciences under Grant ZDKYYQ20200007. 
}

\thanks{Mengzhao Li, Weiyi Sun, Zhijun Liang, Mei Zhao, Yuan Feng, Yunyun Fan, Tianya Wu, Wei Wang, Bo Liu, Xuan Yang, Shuqi Li, Chengjun Yu, Xinhui Huang, Yuekun Heng are with the Institute of High Energy Physics, Chinese Academy of Sciences, Beijing 100049, China. (Corresponding author: Mengzhao Li (mzli@ihep.ac.cn), Mei Zhao (zhaomei@ihep.ac.cn))}

\thanks{Mengzhao Li is also with the China Center of Advanced Science and Technology, Beijing 100190, China.}

\thanks{Weiyi Sun, Yuan Feng, Shuqi Li, Chengjun Yu, Xinhui Huang, Yuekun Heng are also with the School of Physical Sciences, University of Chinese Academy of Sciences, Beijing 100049, China.}

\thanks{Xiaoxu Zhang is with the Nanjing University, Nanjing 210093, China.}
\thanks{Gaobo Xu is with the Institute of Microelectronics, Chinese Academy of Sciences, Beijing 100029, China.}
}

\maketitle

\begin{abstract}
AC-Coupled LGAD (AC-LGAD) is a new 4D detector developed based on the Low Gain Avalanche Diode (LGAD) technology, which can accurately measure the time and spatial information of particles.
The Institute of High Energy Physics (IHEP) designed a large-size AC-LGAD with a pitch of 2000~\SI{}{\micro\metre} and AC pad of 1000~\SI{}{\micro\metre}, and explored the effect of N+ layer dose on the spatial resolution and time resolution.
The spatial resolution varied from 36~\SI{}{\micro\metre} to 16~\SI{}{\micro\metre} depending on N+ dose for a charge corresponding to about 12 minimum ionizing particles. 
The jitter component of the time resolution does not change significantly with different N+ doses, and it is about 15-17 ps measured by laser.
The AC-LGAD with a low N+ dose has a large attenuation factor and better spatial resolution in the central region between pads.
In these specific conditions, large signal attenuation factor and low noise level are beneficial to improve the spatial resolution of the AC-LGAD sensor.

\end{abstract}

\begin{IEEEkeywords}
AC-LGAD, 4D detector, Spatial resolution, Timing resolution
\end{IEEEkeywords}



\section{Introduction}
\label{sec_intuduction}

\IEEEPARstart{L}{ow} Gain Avalanche Diode (LGAD) is a thin N-on-P silicon sensor with a P+ layer (called gain layer) inserted between the N+ layer and the P-type bulk. 
The LGAD sensors can reach a time resolution of about 30 ps~\cite{Nicolo2018,Cartiglia2017,mzli33_2021} thanks to a thin epitaxial layer of 30-50~\SI{}{\micro\metre}, fast rise time and high signal-to-noise ratio.
The LGAD also has good radiation hardness performance and the time resolution can reach 35 ps after \SI{2.5}{\times10^{15}~n_{eq}/\centi\metre^2} irradiation fluence~\cite{G_Kramberger2018,Ugobono2018,Wukw2023}.
One of the important applications of LGAD is the upgrade of the ATLAS and CMS experiments at the High-Luminosity Large Hadron Collider (HL-LHC)~\cite{HGTDtdr2020,MTDtdr2019,HL-LHC}.

\begin{figure}[tbp]
 \begin{center}
\rotatebox{0}{\includegraphics [scale=0.24]{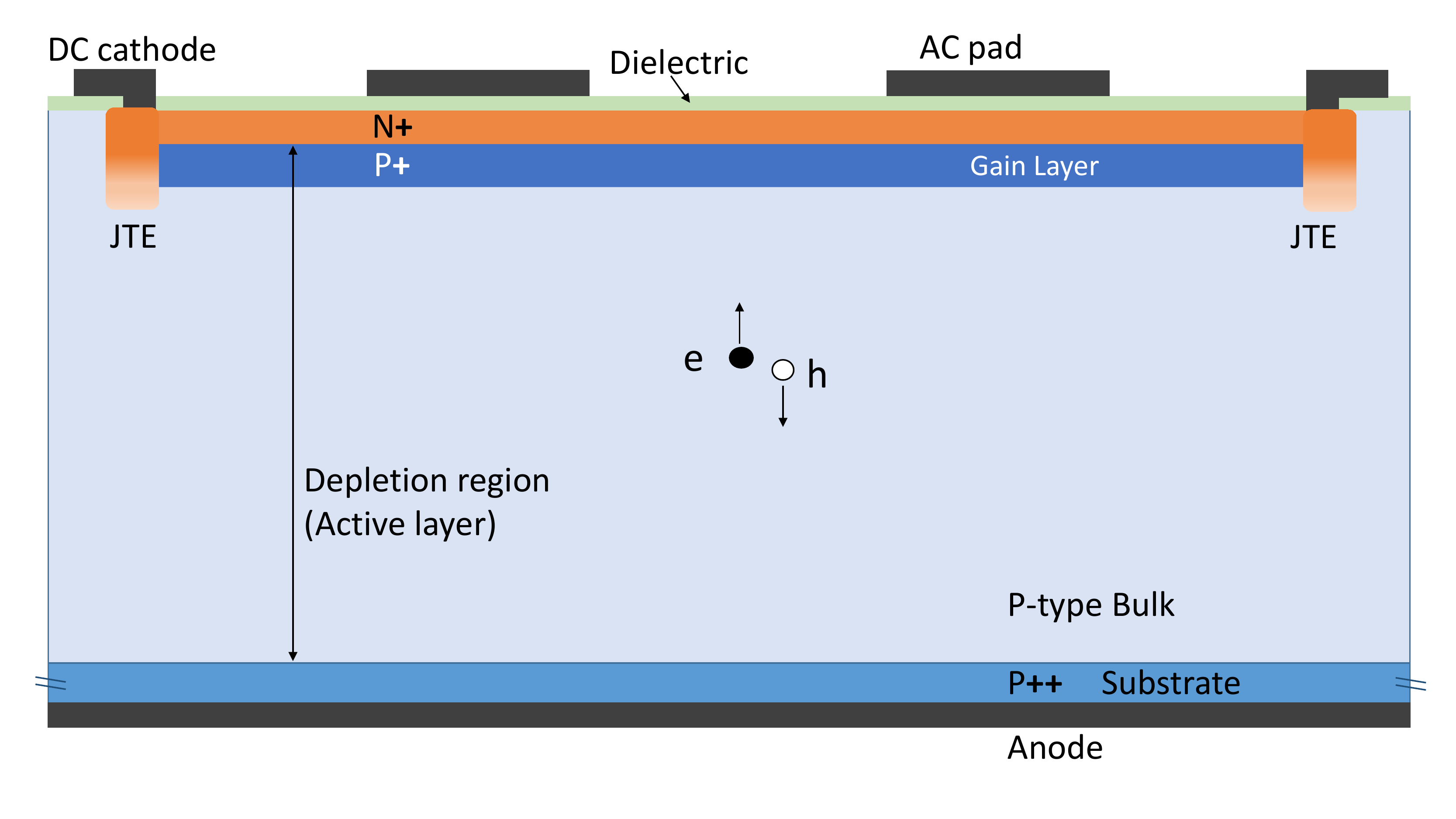}}
\caption{Schematic for the AC-LGAD sensors. The bias voltage is applied between the anode and the DC-cathode, and the Junction Termination Extension (JTE) structure avoids early breakdown at the edge.}
\label{fig:SchematicLGAD}
 \end{center}
\end{figure}

The AC-Coupled LGAD (AC-LGAD) is a new detector based on the LGAD technology. AC-LGADs can be used as a 4D tracker to measure the time and spatial information of particles accurately in the Circular Electron Positron Collider (CEPC)~\cite{CEPC}, the Future Circular Collider (FCC)~\cite{Fcc}, 
and it is planned to be used in the EPIC detector at the Electron-Ion Collider (EIC)~\cite{EIC}. 
The AC-LGAD has uniform and continuous N+ and P+ layers, with a very thin dielectric (\SI{}{SiO_{2}}) grown on the N+ layer, so that the electrodes can be AC-coupled to the N+ layer and segmented, as shown in Fig.~\ref{fig:SchematicLGAD}. 
The continuous P+ layer gives AC-LGAD a 100~$\%$ fill factor. 
When a charged particle or laser pulse hits the AC-LGAD under reverse bias voltage, electron-hole pairs are generated in the epitaxial layer, and the electrons drift to the gain layer and multiply to generate more electrons into the N+ layer.
In such a process, a coupling signal is generated on the AC pad. The signal amplitude is influenced by several parameters, such as pitch size, AC pad size, N+ dose, gain, etc~\cite{Mandurrino2019,Tornago2021}.

Ref.~\cite{Mandurrino2019,Tornago2021,Giacomini2019,RHeller2022} report AC-LGAD sensors produced by FBK, BNL, and HPK with square and strip AC pad electrodes and pitch sizes in the range of 100-550~\SI{}{\micro\metre}.
The large-pitch AC-LGAD is one option for CEPC outer tracker. It reduces the density of the readout channels and the cost by increasing the AC-LGAD pitch size. 
The optimization of the N+ dose is crucial to the development of large-pitch AC-LGAD. However, the effect of the N+ dose has not been extensively studied.
The Institute of High Energy Physics (IHEP) has designed large-pitch (2000~\SI{}{\micro\metre}) AC-LGAD sensors with different N+ doses. 
The first batch of IHEP AC-LGAD sensors is produced by the Institute of Microelectronics (IME) on 8-inch wafers.
In this paper, the performance of the IHEP AC-LGAD with a pitch of 2000~\SI{}{\micro\metre} will be introduced in detail, and the effect of N+ dose on the spatial and time resolution will be reported.

\section{Design parameters of IHEP AC-LGAD sensors}
\label{sec_Properties}

\begin{figure}[!t]
\begin{center}
\includegraphics[scale=0.68]{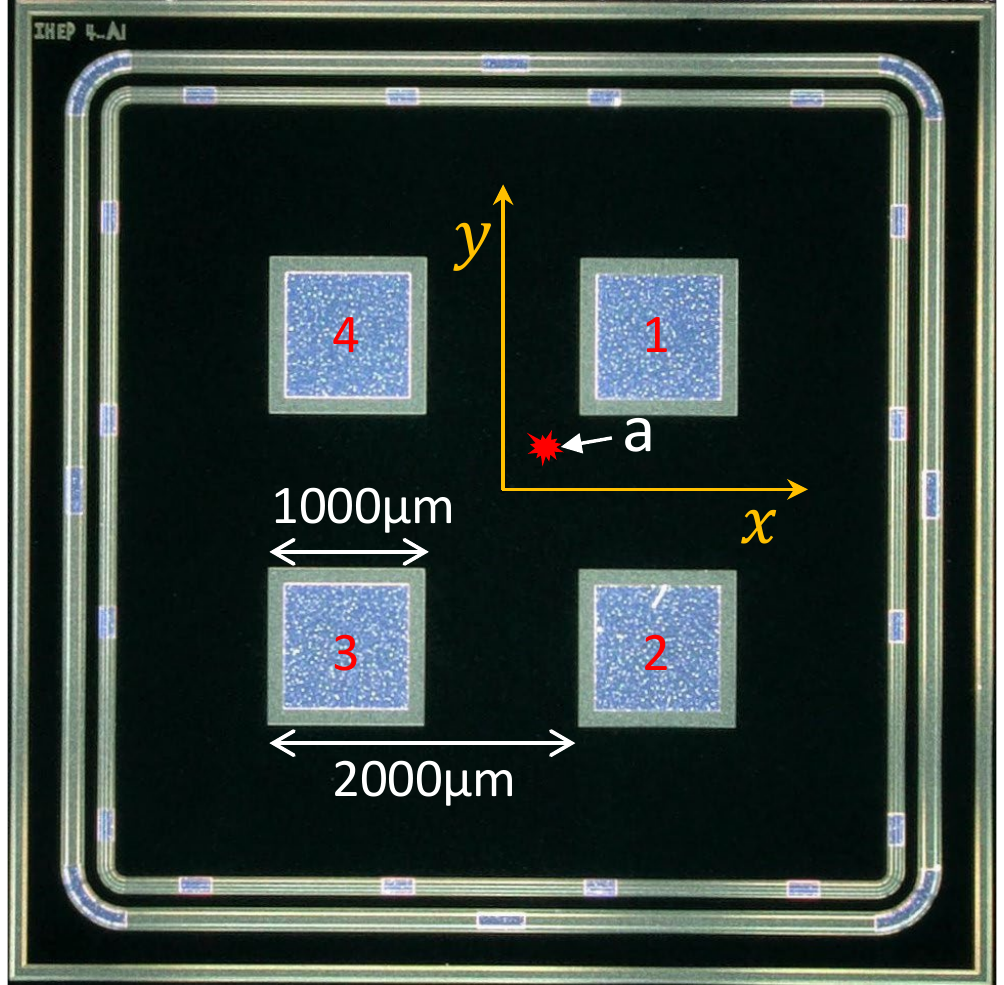} 
\caption{Picture of an IHEP AC-LGAD sensor with pitch size of 2000~\SI{}{\micro\metre} and pad size of 1000~\SI{}{\micro\metre}. The geometric center of the sensor is defined as the coordinate origin in the test, and the mark “a” is a laser hit position.}
\label{fig:picture}
\end{center}
\end{figure}

The IHEP AC-LGADs are fabricated on 8-inch wafers with a 50 \SI{}{\micro\metre} P-type epitaxial layer and a 725 \SI{}{\micro\metre} substrate. 
IHEP AC-LGADs have four square AC pads for AC-coupled signal readout, as shown in Fig.~\ref{fig:picture}. 
The innermost ring is the DC ring (DC-cathode), and the second ring is the guard ring. 
The DC-cathode can be used for DC-coupled signal readout or ground during the test.
The size of the AC pads is 1000~\SI{}{\micro\metre}, and the size of their pitch is 2000~\SI{}{\micro\metre}.

We designed five types of sensors to study the effect of N+ dose on the spatial resolution and time resolution, with five doses of 10.0 P, 5.0 P, 1.0 P, 0.5 P, and 0.2 P. 
Here P is the unit of phosphorus dose defined for the IHEP AC-LGAD.


\section{The experimental setup}
\label{TCT}

\begin{figure}[htbp]
\begin{center}
\subfigure[]{\includegraphics[width=.33\textwidth]{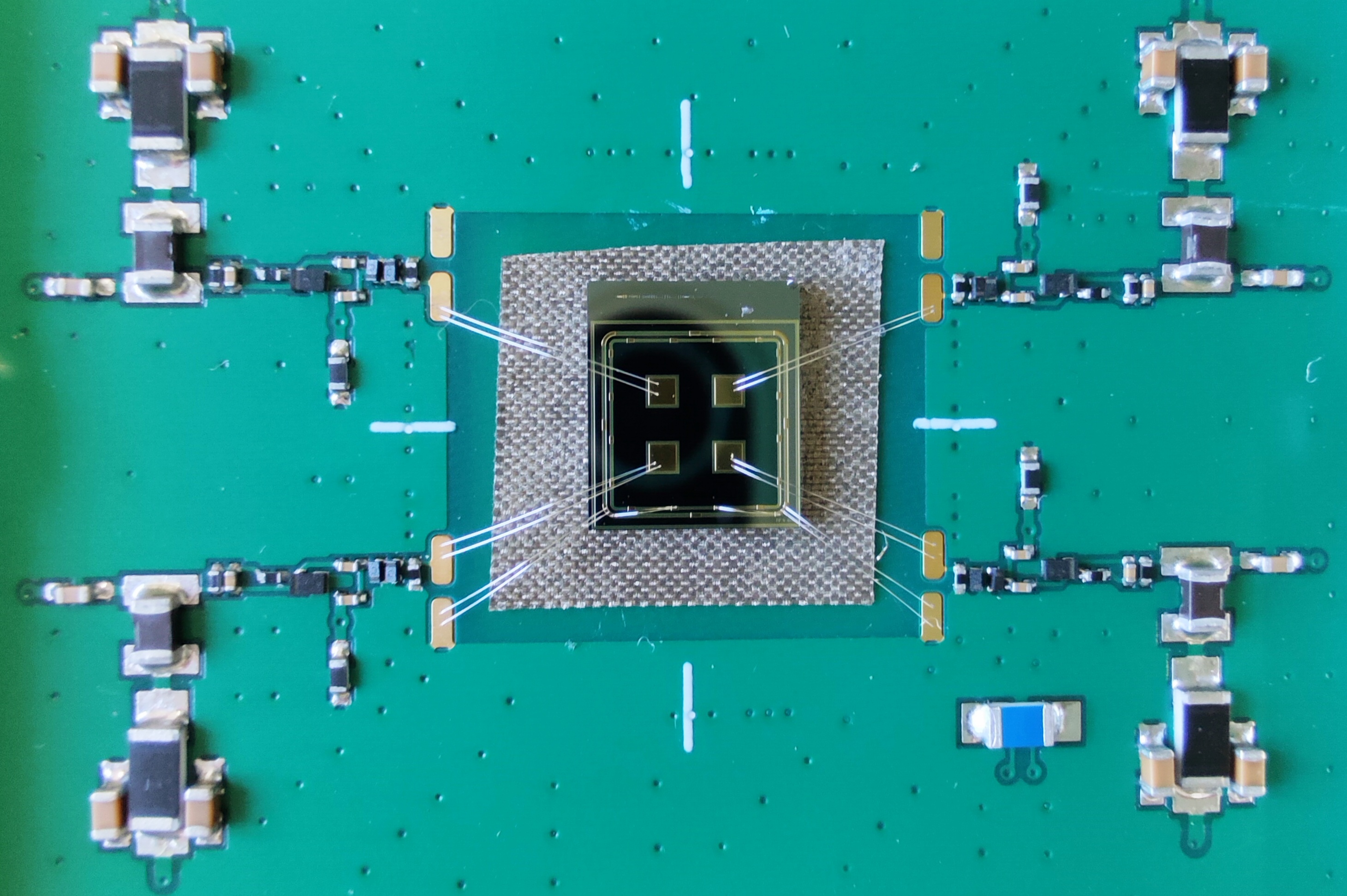} \label{fig:4ChB} }
\subfigure[]{\includegraphics[width=.35\textwidth]{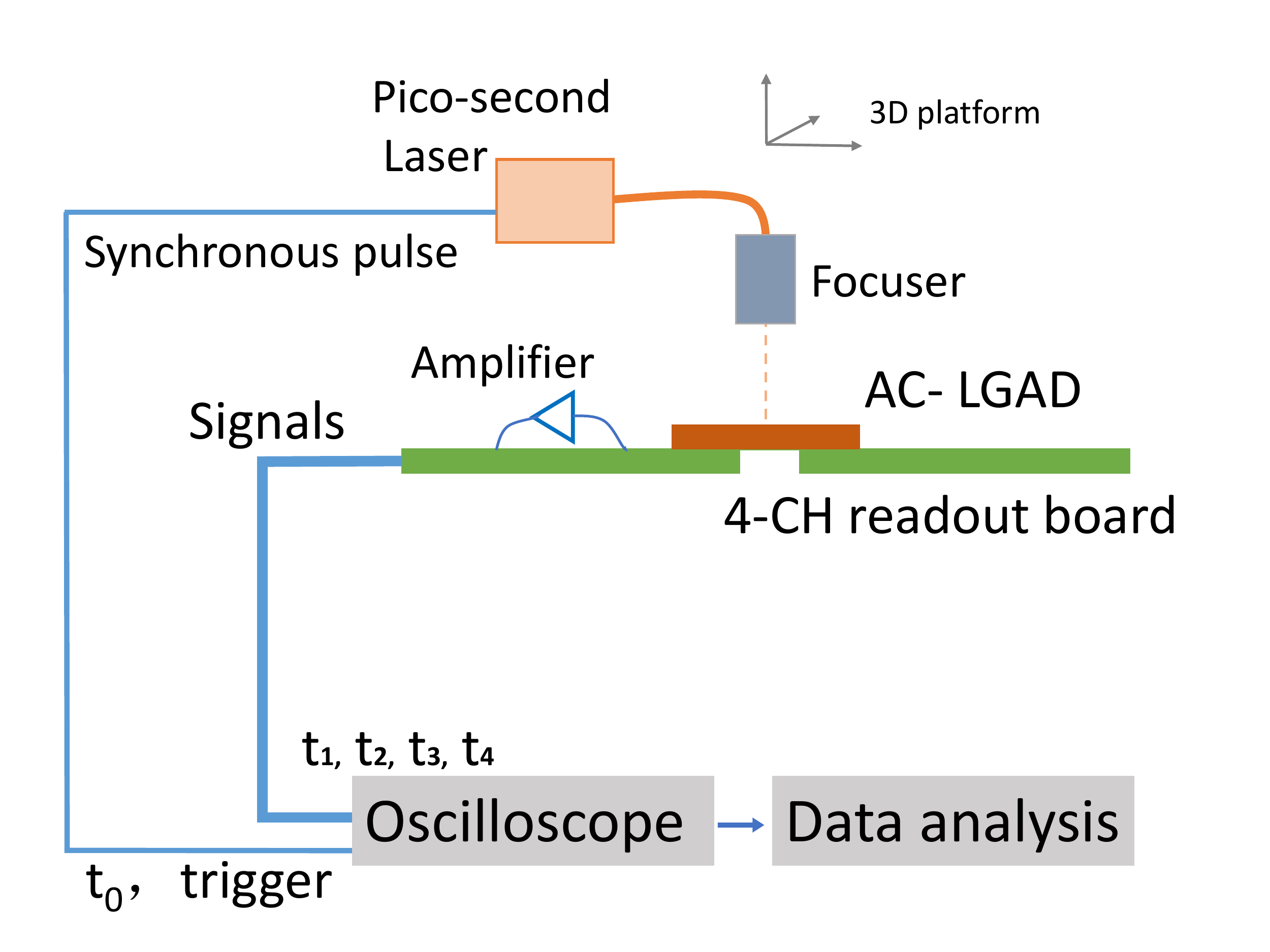} \label{fig:LaserSetup} }
\caption{Laser TCT setup for the AC-LGAD sensors. (a) AC-LGAD is bonded to four channel readout board. (b) The schematic diagram of the TCT platform.}
\label{fig:Laser}
\end{center}
\end{figure}

The IHEP AC-LGAD sensors were tested with a transient current technique (TCT)~\cite{TCT1,TCT2} platform to study the time resolution and spatial resolution. 
The four AC pads are wire bonded to the four channels of the readout board, and the guard ring is grounded, as shown in Fig.~\ref{fig:4ChB}.
The four-channel readout board is designed and manufactured according to the University of California Santa Cruz (UCSC) single-channel readout board~\cite{Cartiglia2017}.
It uses a broadband inverting trans-impedance amplifier of 470~\SI{}{\Omega} for each channel.

\begin{figure}[tbp]
 \begin{center}
\rotatebox{0}{\includegraphics [scale=0.35]{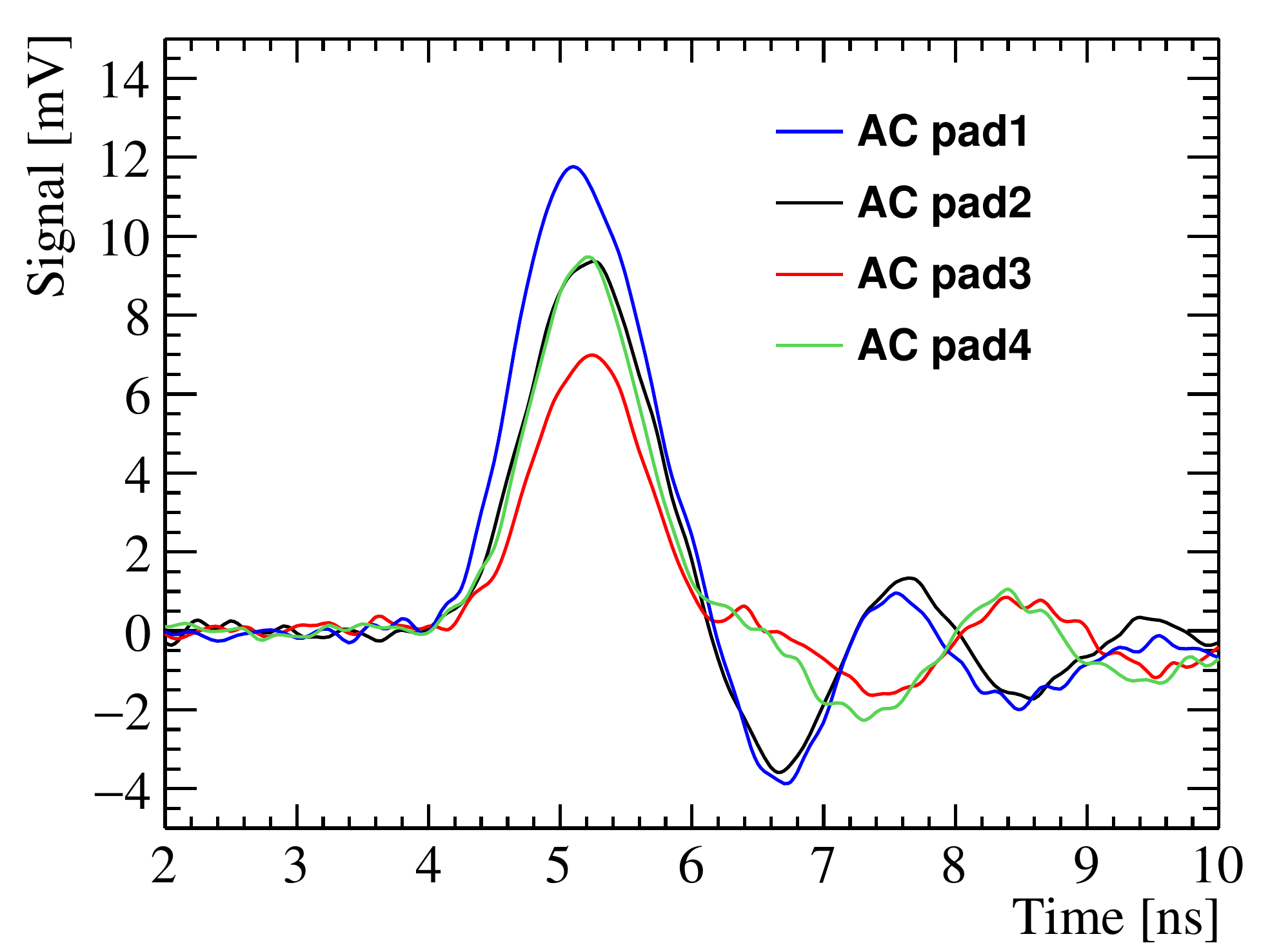}}
\caption{The signals of the four AC pads when the laser hits position “a”.}
\label{fig:waveF}
 \end{center}
\end{figure}

Figure~\ref{fig:LaserSetup} shows the laser TCT experimental setup. The accuracy of the three-dimensional translation platform is about 1~\SI{}{\micro\metre}, the laser wavelength is 1064~nm and the laser spot size is focused to $\sim 10$~\SI{}{\micro\metre} (3 $\sigma$). 
The origin of the coordinates in the test is the geometric center of the sensor (Fig.~\ref{fig:picture}).
Each sensor was tested with a bias voltage corresponding to a gain value of around 30. 
The intensity of the laser pulse is equivalent to about 12 Minimum Ionizing Particles (MIPs).
Figure~\ref{fig:waveF} shows the signals of the four AC pads when the laser hits position “a” (Fig.~\ref{fig:picture}).
The signal amplitude of AC pad 1 closest to “a” is the largest, AC pad 2 and AC pad 4 are the second largest, and AC pad 3 is the smallest.
The signal pulses from four AC pads are recorded by a digital oscilloscope with 2 GHz bandwidth and 40 GS/s sampling rate for offline analysis.

\section{Analysis of experimental results}
\label{}

\subsection{Attenuation of the signal}

\begin{figure}[htbp]
\begin{center}
\subfigure[]{\includegraphics[width=.35\textwidth]{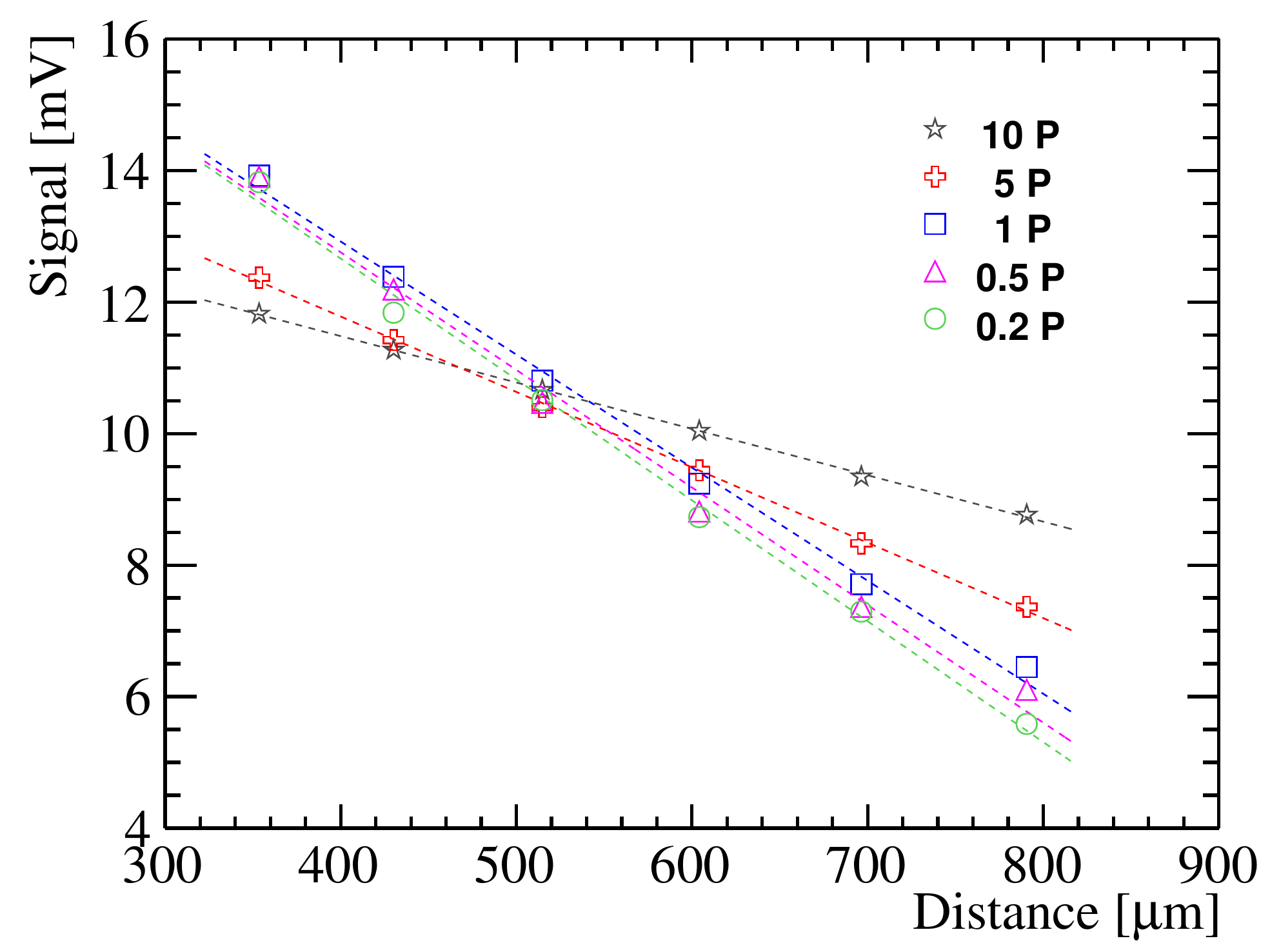} \label{fig:SignalAttenuation} }
\subfigure[]{\includegraphics[width=.37\textwidth]{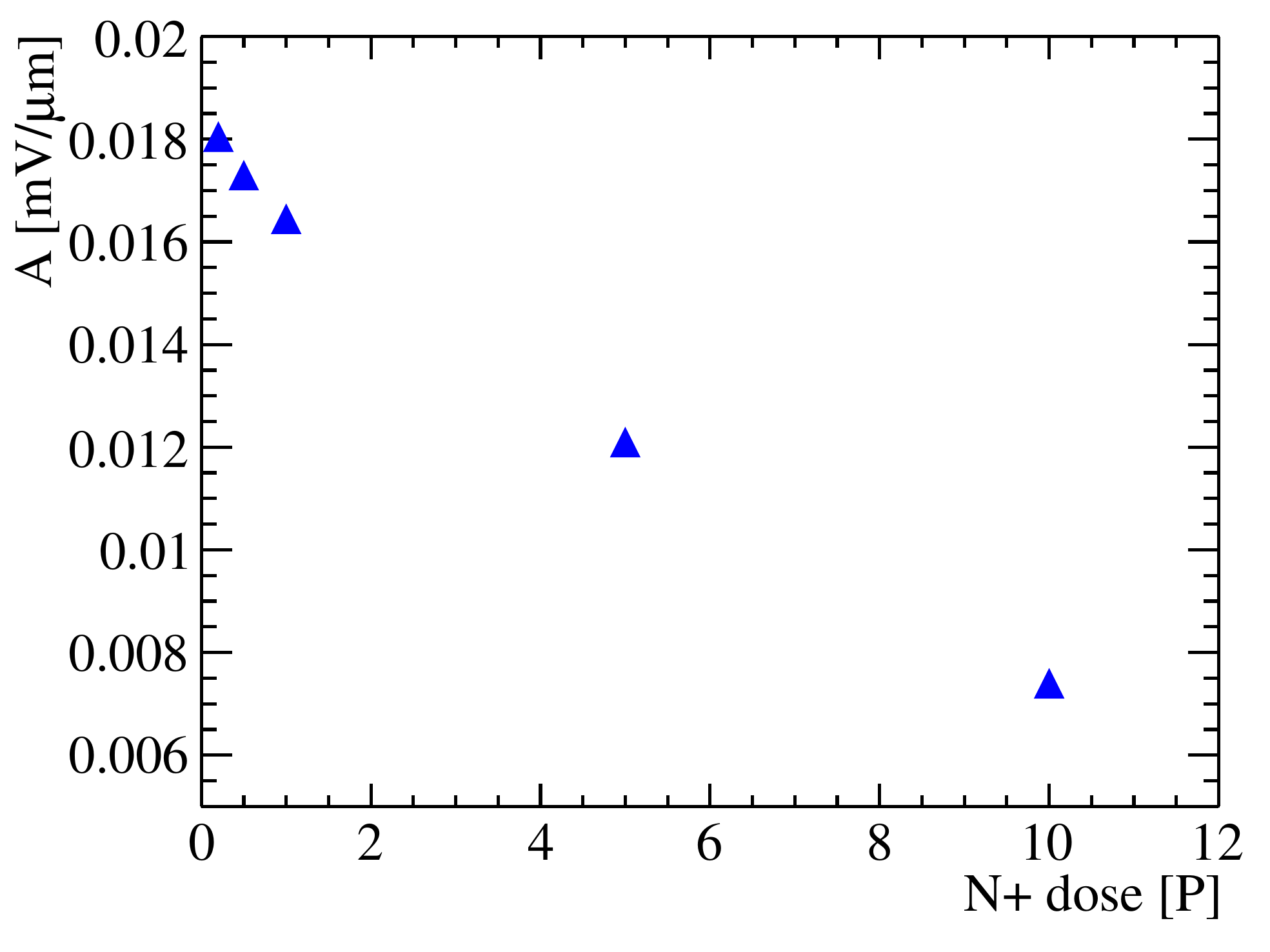} \label{fig:AttenuationFactor} }
\caption{(a) The signal attenuation of AC pad 1 as a function of the distance between the laser spot and the edge of the AC pad 1 for different values of the N+ layer doping. (b) Attenuation factor at different N+ doses.}
\label{fig:SignalA}
\end{center}
\end{figure}

The signal amplitude of an AC pad usually decreases with the distance between the laser spot and the AC pad~\cite{Tornago2021,Giacomini2019}.
Figure~\ref{fig:SignalAttenuation} shows that the signal amplitude changes with the distance between the laser spot and the edge of the AC pad. 
The signal amplitude decreases as the laser spot distance increases. 
The decrease of the signal amplitude in the central region between the two AC pads is very close to linear attenuation, not shown is the non-linearity trend in the region close to the electrode. 
Linear fitting is performed for signal amplitude and laser spot distance, and the slope is defined as the signal attenuation factor $A$.
The attenuation factor $A$ of the IHEP AC-LGAD sensors is shown in Fig.~\ref{fig:AttenuationFactor}. 
The attenuation factor increases for decreasing N+ doses.
The N+ layer with a low dose has high resistivity and a large signal attenuation factor, which also causes the signal to be more sensitive to the change of laser position.

\subsection{Position reconstruction and spatial resolution}

\begin{figure}[tbp]
 \begin{center}
\rotatebox{0}{\includegraphics [scale=0.4]{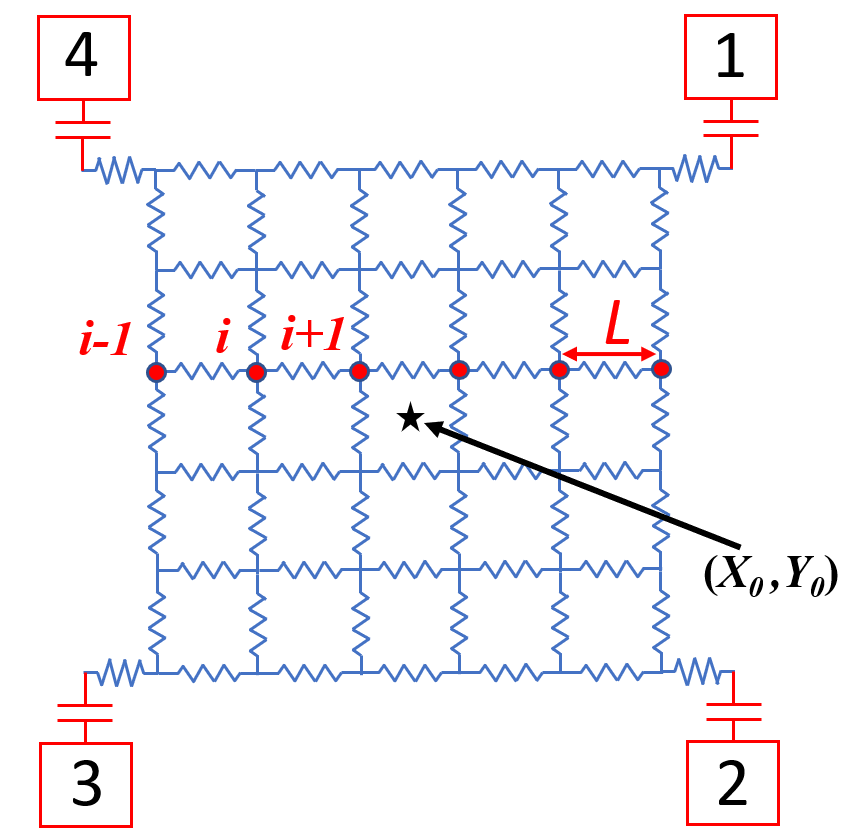}}
\caption{The discretized positioning circuit representation of an AC-LGAD~\cite{Tornago2021}.}
\label{fig:DPC}
 \end{center}
\end{figure}

In the process of the signal induced by a laser pulse, the sensor can be considered as a discretized positioning circuit (DPC)~\cite{Tornago2021}, as shown in Fig.~\ref{fig:DPC}. The N+ layer is equivalent to a resistive array, and the induced charges on the four AC pads $1$, $2$, $3$, and $4$ are $q_{1}$, $q_{2}$, $q_{3}$, and $q_{4}$. The hit position $(X, Y)$ of the laser or particle is determined by using the charge or amplitude imbalance between AC pads along the X and Y directions:

\begin{equation} \label{eq:DPC}
    \begin{cases}
    X = X_{0 } + k_{x}m \\
    Y = Y_{0 }  + k_{y}n \\
    m = \frac{ q_{1} + q_{2} - q_{3} - q_{4} }{ q_{1} + q_{2} + q_{3} + q_{4} }\\
    n = \frac{ q_{1} + q_{4} - q_{2} - q_{3} }{ q_{1} + q_{2} + q_{3} + q_{4} }
    \end{cases}
\end{equation}

\begin{equation} \label{eq:DPCKxKy}
    \begin{cases}
    k_{x} = L \frac{ \sum (m_{i+1} - m_{i}) }{ \sum (m_{i+1} - m_{i})^{2} } \\
    k_{y} = L \frac{ \sum (n_{i+1} - n_{i}) }{ \sum (n_{i+1} - n_{i})^{2} } \\
    \end{cases}
\end{equation}
where $k_{x}$ and $k_{y}$ are correction factors along X direction and Y direction respectively, and $m_{i}$ and $n_{i}$ are the $m$ and $n$ values for the ith test postion.
In order to obtain the correction factors $k_{x}$ and $k_{y}$ according to equation~\ref{eq:DPCKxKy}, the method is as follows.
First, move the laser along the X direction with a step size $L$, calculate $m_{i}$ for each position, and bring them into equation~\ref{eq:DPCKxKy}.
Similarly, $k_{y}$ can also be obtained by a series of equidistant test positions along the Y direction.
The reconstruction method mentioned above based on the DPC model is used for the central region of AC-LGAD, where the signal is linearly attenuated.

The TCT system is used to complete a \SI{6}{\times6} laser test array in the central region of the AC-LGAD, and the step size of the laser is~\SI{100}{\micro\metre}.
Each position in the \SI{6}{\times6} array has 1000 events recorded by the oscilloscope.
The coordinates of the laser hit position will be reconstructed by equation~\ref{eq:DPC}.
The spatial resolution of each sensor is evaluated as the Root Mean Square (RMS) of the difference between the reconstructed coordinates and the laser spot coordinates.

\begin{figure}[tbp]
 \begin{center}
\rotatebox{0}{\includegraphics [scale=0.35]{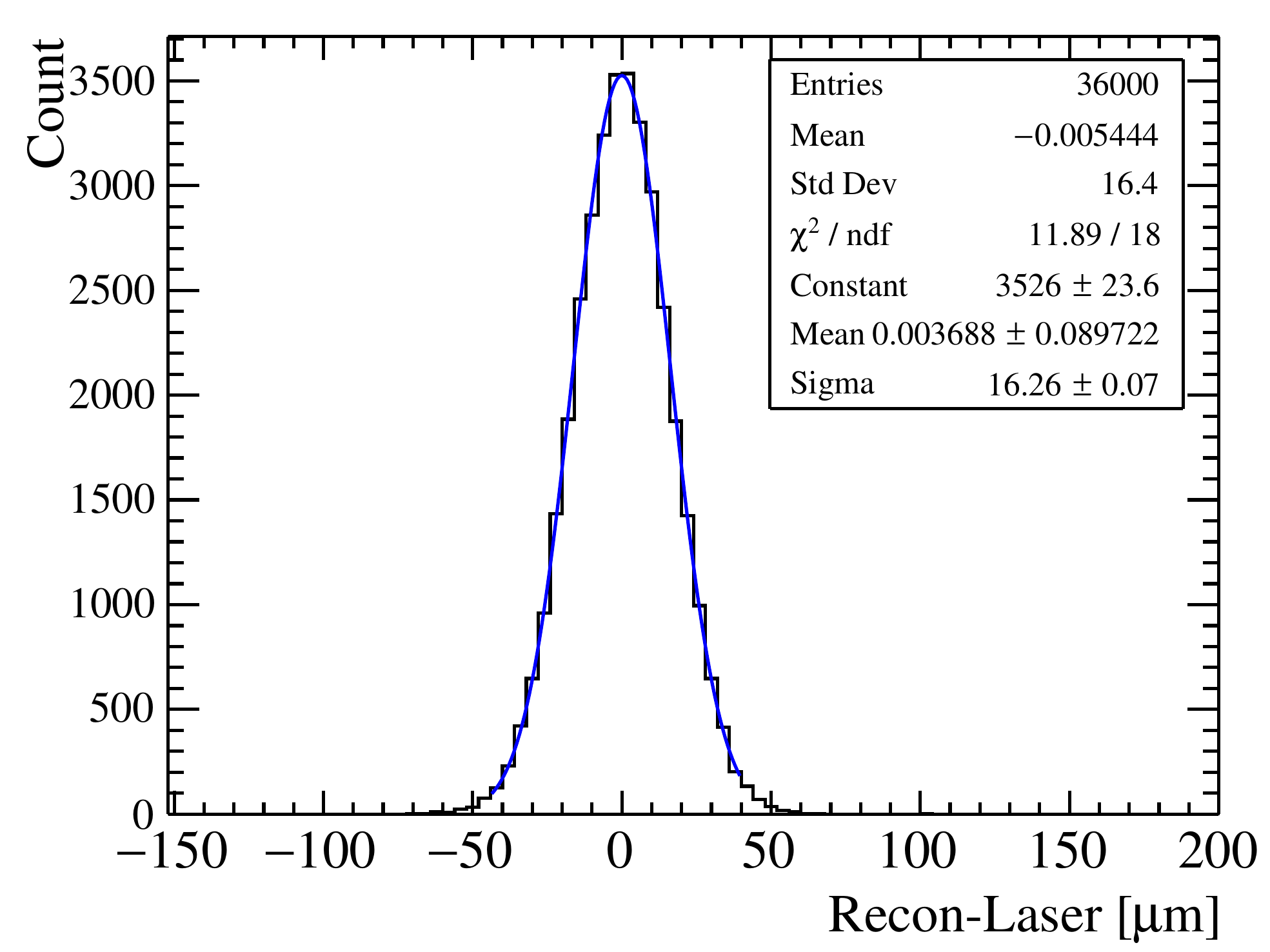}}
\caption{The distribution of the difference between the reconstructed coordinates and the laser spot coordinates of the sensor with dose 0.2 P, and fit with a Gaussian function.}
\label{fig:posX}
 \end{center}
\end{figure}

\begin{figure}[tbp]
 \begin{center}
\rotatebox{0}{\includegraphics [scale=0.35]{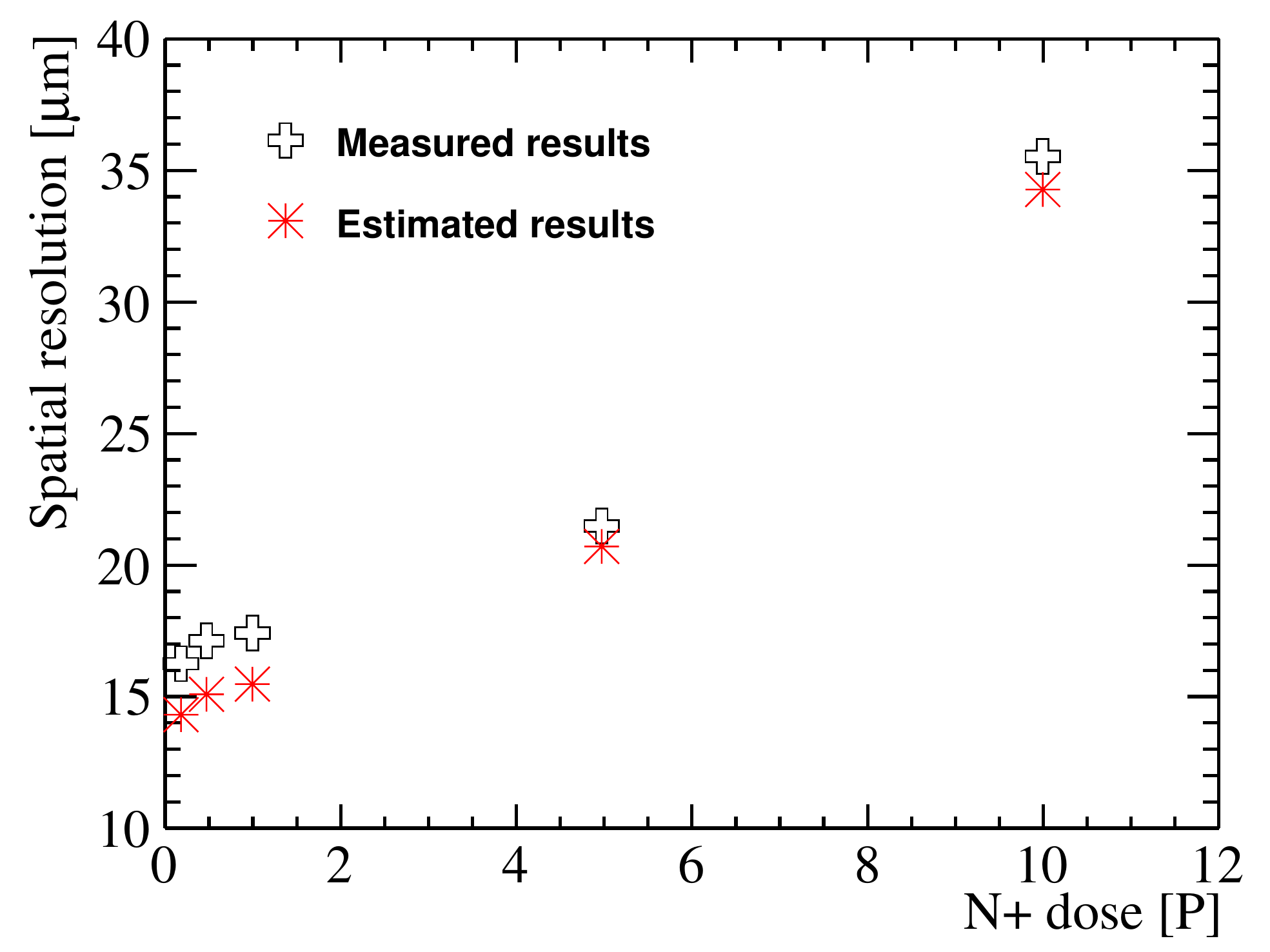}}
\caption{The directly-measured spatial resolution (black marks) and estimated spatial resolution (red marks) at different N+ doses.}
\label{fig:spatial}
 \end{center}
\end{figure}

\begin{figure}[tbp]
 \begin{center}
\rotatebox{0}{\includegraphics [scale=0.35]{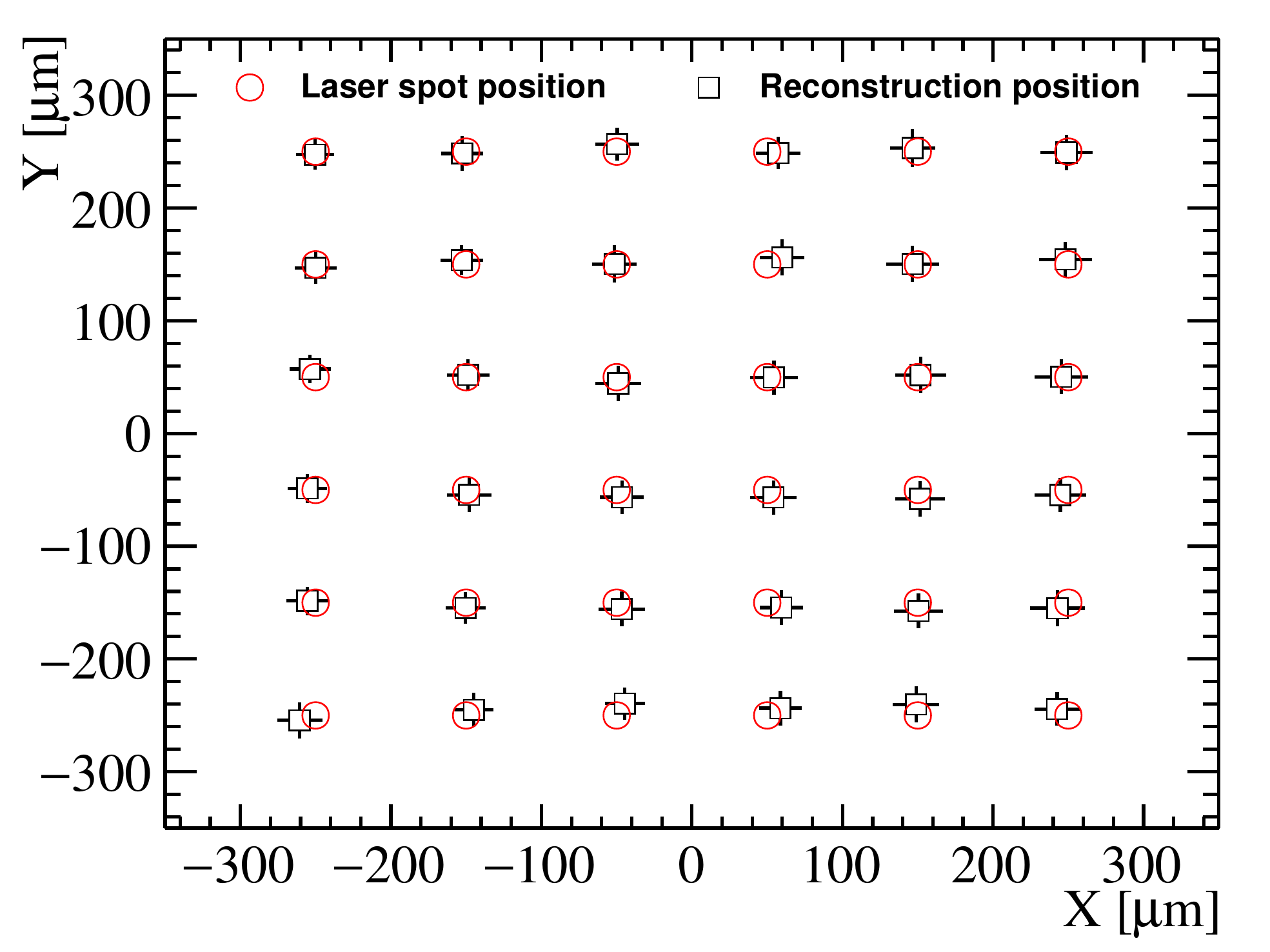}}
\caption{The reconstructed position of the \SI{6}{\times6} laser array of the sensor with dose 0.2 P (black squares) and the laser spot position as measured by the TCT stage (red circles).}
\label{fig:PointRec}
 \end{center}
\end{figure}

Figure~\ref{fig:posX} shows the distribution of the difference between the reconstructed coordinates and the laser spot coordinates of the sensor with dose 0.2 P, including \SI{}{6\times6\times1000} events, and the RMS value is \SI{16}{\micro\metre}.
Therefore, the spatial resolution of the test region of the sensor with a dose of 0.2 P is~\SI{16}{\micro\metre}, which is also called the directly-measured spatial resolution in the following.
The directly-measured spatial resolution as a function of N+ dose is shown in Fig.~\ref{fig:spatial}.
As the N+ dose is reduced from 10 P to 0.2 P, the spatial resolution improves from~\SI{36}{\micro\metre} to~\SI{16}{\micro\metre}.
Figure~\ref{fig:PointRec} shows the reconstruction results of the \SI{6}{\times6} laser test array of the sensor with dose 0.2 P. 
The reconstructed array agrees well with the laser test array, with only slight deviations.

The sensitivity of the AC-LGAD sensor to the movement of the laser hit position depends on the change of AC pad signal amplitude and the noise level (sensors and electronics). When the amplitude change of the signal is less than the noise, the movement of the laser hit position cannot be recognized in the used setup.
The amplitude change of the signal is proportional to the attenuation factor.
Therefore, we we empirically assume that the spatial resolution is proportional to the signal attenuation factor ($A$) and inversely proportional to the noise RMS value ($N$):
\begin{equation}
\sigma_{spatial} \approx \frac{ N }{A }
\end{equation}
The noise $N$ of the five types of IHEP AC-LGAD sensors in the TCT test was very similar, all around 0.25 mV. 
As shown in Fig.~\ref{fig:spatial}, the spatial resolution estimated by the attenuation factor $A$ and noise $N$ agrees with the directly measured results.
Therefore, AC-LGAD spatial resolution can also be evaluated according to signal attenuation factor and noise level.
Reducing the noise level and appropriately increasing the signal attenuation factor are effective methods to improve spatial resolution in the central region of the sensor between the AC pads. 

Decreasing the dose or concentration of phosphorus increases the resistivity of the N+ layer and the attenuation factor of the AC pad signal.
However, an optimal N+ dose needs further studies, because an excessive reduction of the N+ dose will increase the attenuation factor, which in turn may reduce the signal amplitudes at large distances from the electrodes, and the signal-to-noise ratio.

\subsection{The time resolution}

The arrival time of particle or laser ($t_\mathrm{arrived}$) is defined as the mean value of the cross-threshold time of four AC pad signals:
\begin{equation}\label{eq:timing1}
t_{\mathrm{arrived}} = ({t_{\mathrm{1}} + t_{\mathrm{2}} + t_{\mathrm{3}} + t_{\mathrm{4}}}) / 4 
\end{equation}
where $t_{\mathrm{1}}$, $t_{\mathrm{2}}$, $t_{\mathrm{3}}$, $t_{\mathrm{4}}$ are the cross-threshold time of four AC pads obtained according to the constant fraction discriminator (CFD) method.
In this experimental setup, the spread of arrival time is mainly composed of the time resolution of AC-LGAD ($\sigma_\mathrm{ACtime}$) and the jitter of the trigger $t_\mathrm{0}$ ($\sigma_\mathrm{t_\mathrm{0}}$):
\begin{equation}
\sigma^2_{({t_{\mathrm{1}} + t_{\mathrm{2}} + t_{\mathrm{3}} + t_{\mathrm{4}}}) / 4} =  \sigma^2_{ACtime} + \sigma^2_{ t_\mathrm{0}}
\end{equation}
To avoid the jitter of $t_\mathrm{0}$, $({t_{\mathrm{1}} + t_{\mathrm{2}} - t_{\mathrm{3}} - t_{\mathrm{4}}}) / 4$ is used to calculate the time resolution of AC-LGAD sensors:
\begin{equation}
\sigma_{ACtime} =  \sigma_{({t_{\mathrm{1}} + t_{\mathrm{2}} - t_{\mathrm{3}} - t_{\mathrm{4}}}) / 4}.
\end{equation}

\begin{figure}[!htbp]
\centering
\includegraphics[scale=0.35]{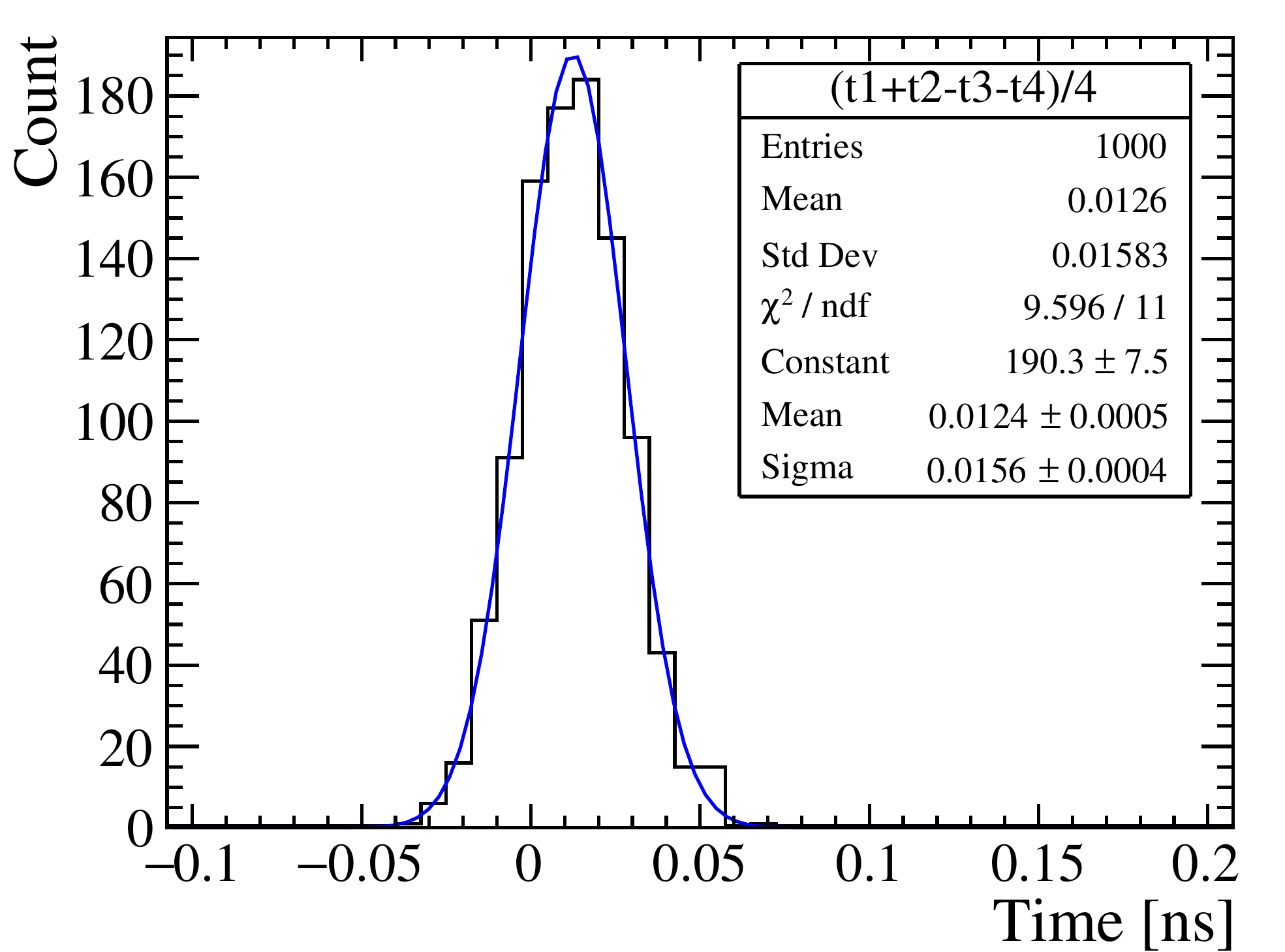} 
\caption{Distribution of $({t_{\mathrm{1}} + t_{\mathrm{2}} - t_{\mathrm{3}} - t_{\mathrm{4}}}) / 4$ at one position in the \SI{6}{\times6} laser test array.}
\label{fig:timing_one_point}
\end{figure}

\begin{figure}[!htbp]
\centering
\includegraphics[scale=0.35]{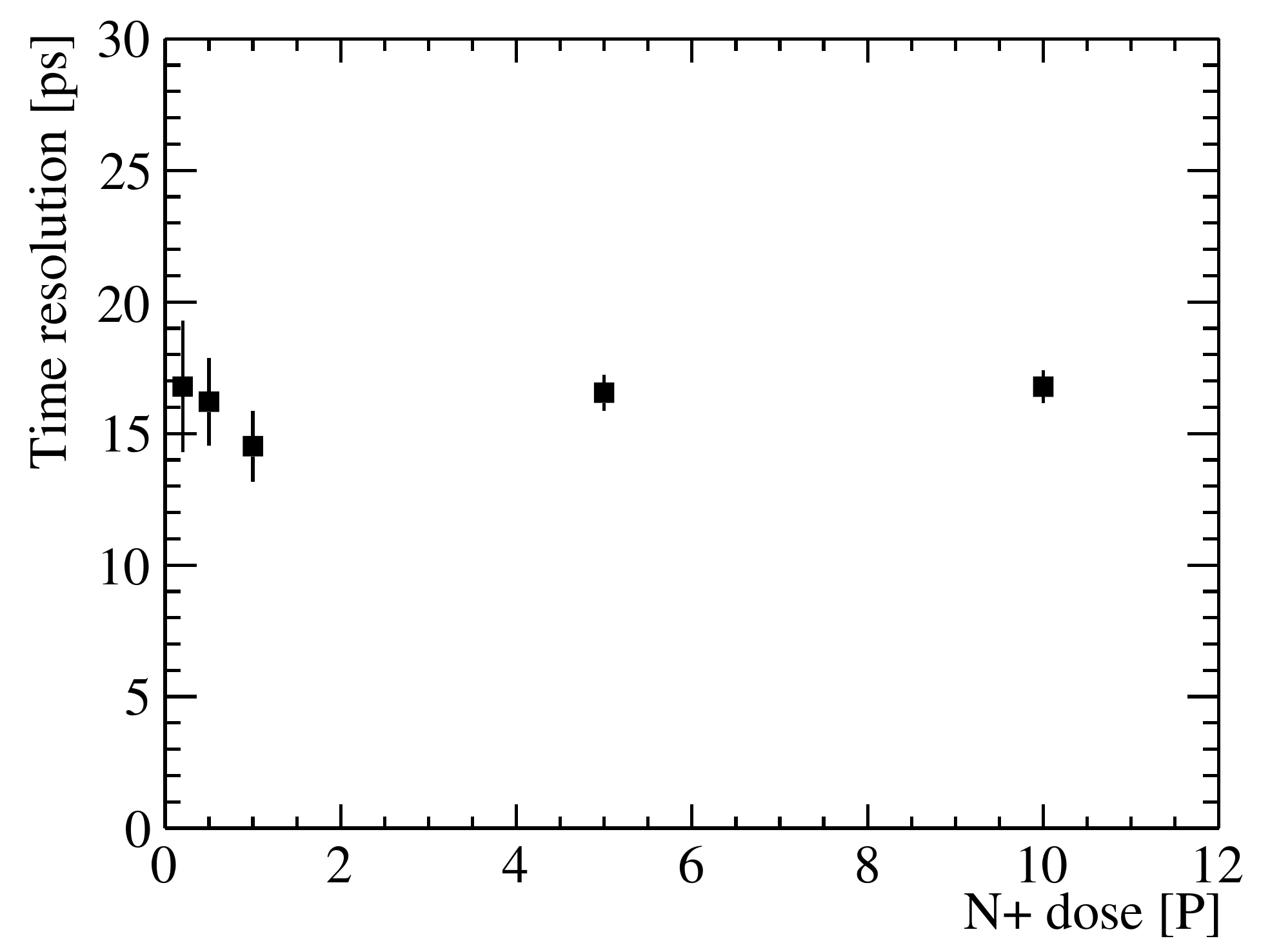} 
\caption{The jitter component of the time resolution with different N+ doses.}
\label{fig:timing2}
\end{figure}

Figure~\ref{fig:timing_one_point} shows the distribution of $({t_{\mathrm{1}} + t_{\mathrm{2}} - t_{\mathrm{3}} - t_{\mathrm{4}}}) / 4$, with a time resolution of 15.6 ps. The time resolution here is based on laser tests and includes only the jitter component~\cite{mzli33_2021,Nicolo2018}, which is evaluated by the mean value of the jitter component of 36 test positions.
Figure~\ref{fig:timing2} shows the jitter component of the time resolution of AC-LGAD with different N+ doses. 
The jitter component of the time resolution varies slightly, about 15-17 ps with different N+ doses.

\section{Conclusion}
\label{sec_Conclusion}

The AC-LGAD sensor with a large-pitch size has a lower readout electronic density, which can be applied to CEPC and FCC detectors in the future.
The first batch of IHEP AC-LGAD sensors was designed by IHEP and fabricated by IME with a 2000~\SI{}{\micro\metre} pitch size and 1000~\SI{}{\micro\metre} AC pad size.
Five types of AC-LGAD sensors with different N+ doses (10.0 P, 5.0 P, 1.0 P, 0.5 P, and 0.2 P) were designed to study the effect of N+ dose on the spatial and time resolution.
We tested the central regions of these sensors using a laser TCT platform.
The jitter component of the time resolution does not change significantly with different N+ doses, and it is about 15-17~ps measured with the laser, when its intensity corresponds to the charge deposited by about 12 MIPs.
With such large signals and in the central region of the sensor, between the AC pads, the sensor with a low N+ dose is found to have small spatial resolution in the range of 16-36~\SI{}{\micro\metre}, for doses in the range of 0.2 P to 10.0 P.
Reducing the dose or concentration of phosphorus increases the N+ layer resistivity and the signal attenuation factor, which can improve the spatial resolution.
The spatial resolution is positively correlated with the signal attenuation factor $A$ and negatively correlated with the noise $N$.
In the sensor region under test, reducing the noise and increasing the signal attenuation factor appropriately is an effective way to improve spatial resolution.



\appendices

\section*{References}


\def\refname{\vadjust{\vspace*{-1em}}} 








\end{document}